# Dense blocks of energetic ions driven by multi-petawatt lasers

S. M. Weng[1,2,*], M. Liu[1,2], Z. M. Sheng[1,2,3,*], M. Murakami[4], M. Chen[1,2], L. L. Yu[1,2], and J. Zhang[1,2]

[1]Key Laboratory for Laser Plasmas (Ministry of Education), Department of Physics and Astronomy, Shanghai Jiao Tong University, Shanghai 200240, China

[2]Collaborative Innovation Center of IFSA (CICIFSA), Shanghai Jiao Tong University, Shanghai 200240, China

[3]SUPA, Department of Physics, University of Strathclyde, Glasgow G4 0NG, UK

[4]Institute of Laser Engineering, Osaka University, Osaka 565-0871, Japan

Correspondence and requests for materials should be addressed to S.M.W. (wengsuming@sjtu.edu.cn) or to Z.M.S. (zhengming.sheng@strath.ac.uk).

**Abstract**

**Laser-driven ion accelerators have the advantages of compact size, high density, and short bunch duration over conventional accelerators. Nevertheless, it is still challenging to simultaneously enhance the yield and quality of laser-driven ion beams for practical applications. Here we propose a scheme to address this challenge via the use of emerging multi-petawatt lasers and a density-modulated target. The density-modulated target permits its ions to be uniformly accelerated as a dense block by laser radiation pressure. In addition, the beam quality of the accelerated ions is remarkably improved by embedding the target in a thick enough substrate, which suppresses hot electron refluxing and thus alleviates plasma heating. Particle-in-cell simulations demonstrate that almost all ions in a solid-density plasma of a few microns can be uniformly accelerated to about 25% of the speed of light by a laser pulse at an intensity around $10^{22}$ W/cm$^2$. The resulting dense block of energetic ions may drive fusion ignition and more generally create matter with unprecedented high energy density.**

Inertial confined fusion, driven either by ion beams or by laser pulses, has been extensively studied over 3-4 decades due to its potential in settling mankind's energy problem. However, the ignition of inertial confined fusion remains highly challenging because it requires the fast heating of a compressed fusion target to a sufficiently high temperature[1,2]. Currently, ion beams from conventional accelerators cannot offer sufficient instantaneous beam intensity for such fast heating, while laser pulses cannot deposit their energy directly in compressed fusion targets. On the other hand, laser-driven acceleration of high-energy ion beams has received a growing interest with the development of ultra-short high power lasers. Compared with conventional accelerators, laser-driven ion accelerators have various advantages, including higher ion density, shorter bunch duration, and more compact size[3-5]. These advantages may promise novel ion sources in specialized applications, such as fast ignition of laser fusion[6,7], high energy density physics[8], radiotherapy[9] and radiography[10].

Over the past two decades, a variety of schemes have been proposed for laser-driven ion acceleration[4,5]. Most of these schemes can be categorized as one or a combination of the following mechanisms: target normal sheath acceleration (TNSA)[11], break-out-afterburner (BOA) acceleration[12,13], Coulomb explosion[14], collisionless shock acceleration[15], and radiation pressure acceleration (RPA)[16-18]. In RPA, the light pressure of an intense circularly-polarized laser pulse can push a substantial number of electrons forward, resulting in a strong charge-separation field for ion acceleration. In particular, there are two distinct regimes of RPA according to the thicknesses of irradiated targets[19,20]. In the "light sail" regime of RPA[21-24], the ions of an ultrathin foil can be accelerated continuously to ultrahigh energies. By contrast, the ions of a relatively thick target can be accelerated layer by layer from the front surface in the so-called "hole-boring" regime of RPA[25,26].

As the hole-boring RPA could work with a relatively thick and dense target, it has the potential to generate an energetic ion beam with both a high flux and a high fluence[19]. Consequently, the hole-boring RPA may provide a competitive ion driver for fast ion ignition of laser fusion[6] and even for heavy-ion fusion[27]. To achieve fast ion ignition, the energy flux and fluence of the ion beam should be as high as $10^{20}$ W/cm$^2$ and GJ/cm$^2$ (~10 kJ in a duration of ~20 ps and a volume of linear dimension ~20 μm)[28], respectively. Such a high-flux and high-fluence ion beam still challenges traditional radio-frequency or induction accelerators. With regard to the hole-boring RPA, however, there are also a few key challenges. For instance, it is still hard to control the acceleration stability and the beam quality in the hole-boring RPA.

Moreover, the target is usually assumed to be cold enough, so other ion acceleration mechanisms could be ignored in the hole-boring RPA. In fact, this assumption is usually false because the electrons of the target are easily heated as long as the ions are accelerated. To solve these issues, a few ingenious designs have been proposed in the hole-boring RPA, such as using sandwich targets[29] or elliptically polarized laser pulses[30,31]. Particularly, the high-quality hole-boring RPA has been observed experimentally via an intense infrared $CO_2$ laser interaction with a gaseous hydrogen target[32]. In marked contrast to solid targets, gas targets open the way to hole-boring RPA at a greatly reduced intensity and are also suitable for high-repetition-rate operation. Recently, it has also been proposed to shape the profile of a laser pulse for an efficient and high-quality hole-boring RPA in an inhomogeneous plasma[33]. Unfortunately, these proposals are often somewhat single-minded. That is to say, the quality of the ion beam is improved at the cost of decreasing the energetic ion yield. Meanwhile, the precise shaping of the profile of an ultra-short ultra-intense laser pulse is still challenging with current laser techniques.

In this article, we propose a route towards the hole-boring RPA of ion beams with high yield and high quality. Firstly, the density profile of a solid target is modulated according to the intensity profile of an incident laser pulse for efficient laser-target coupling. With the help of the pulsed laser deposition technique[34], or the layer-by-layer nanoarchitectonics[35], it might be more practical to modulate the target density profile than the laser intensity profile. Secondly, the density modulated (DM) target is half embedded in a substrate thick enough to keep the DM target cold enough for the high-quality hole-boring RPA. Using particle-in-cell (PIC) simulations, we verify that such a DM target can be uniformly accelerated as a whole by a multi-petawatt Gaussian laser pulse. Particularly, the accelerated ions group together in a highly dense block, and they have the feature of high density, narrow energy spread and small spatial divergence. Our results demonstrate that the ion yield together with the beam quality can be dramatically enhanced by target density modulation and structure optimization in laser-driven ion acceleration.

## Results

**Theoretical and numerical modelling.** By using a quasi-stationary laser piston model, the hole-boring velocity $v_b$ and the peak energy of ions $\varepsilon_i$ can be obtained from the momentum flux balance as[25,26]

$$\frac{v_b}{c} = \frac{\sqrt{\Xi}}{1+\sqrt{\Xi}}, \quad \frac{\varepsilon_i}{m_i c^2} = \frac{2\Xi}{1+2\sqrt{\Xi}}, \quad (1)$$

where the "pistoning" parameter $\Xi \equiv I/\rho c^3$ is defined by the laser intensity $I$, the mass density of target $\rho$, and the speed of light $c$. For simplicity, we introduce the dimensionless amplitude of laser electric field $a \equiv |e\mathbf{E}/m_e \omega c|$, where $e$ and $m_e$ are the electron charge and mass, and $\mathbf{E}$ and $\omega$ are the laser electric field and frequency. For a circularly-polarized pulse $a = (I/m_e n_c c^3)^{1/2} = [I\lambda^2/(2.74 \times 10^{18}\ \text{Wcm}^{-2}\mu m^2)]^{1/2}$, where the critical density $n_c \equiv m_e \epsilon_0 \omega^2/e^2$ and $\lambda$ is the laser wavelength. Then we can rewrite the pistoning parameter of hole-boring as $\Xi \equiv a^2 m_e n_c / \sum_i (m_i n_i)$, the sum is over all ion species $i$, and $m_i$ and $n_i$ are the ion mass and density, respectively.

Equation (1) indicates that a uniform hole-boring velocity, i.e., a constant parameter $\Xi \equiv \Xi_0$, is essential for the generation of a monoenergetic ion beam[33]. Assuming the laser propagates along the $x$-axis, the target density should be modulated as $\rho(x,r) = I(t,r)/c^3 \Xi_0$ to yield a constant $\Xi_0$, where $x = c\sqrt{\Xi_0} t$ and $r$ is the radial distance from the $x$-axis. For a Gaussian laser pulse, we have $I(t,r) = I_0 \exp(-2.77 t^2/\tau_L^2)\exp(-r^2/\sigma^2)$, where $I_0$, $\tau_L$, and σ are the peak intensity, full width at half maximum (FWHM) duration, and spot size, respectively. Correspondingly, the density profile of the DM target can be specialized as

$$\rho(x,r) = \rho_0 \exp(-2.77\frac{x^2}{L_x^2})\exp(-\frac{r^2}{\sigma^2}), \quad (2)$$

where $\rho_0 = I_0/c^3\Xi_0$ and $L_x = c\sqrt{\Xi_0}\tau_L$ are the peak density and FWHM dimension in $x$ direction, respectively.

Figure 1 depicts the schematic of a uniform hole-boring RPA using a DM target. Given an incident laser pulse, the density profile of the DM target is modulated according to Eq. (2). By dosing so, the ions of the target can be accelerated layer by layer to achieve a roughly uniform speed. In addition, a uniform dense substrate of enough thickness is employed as a heat conductor to cool the DM target, which we will discuss later in detail.

To verify the uniform and high-quality hole-boring RPA using our scheme, the interactions of intense laser pulses with different target configurations are modelled using particle-in-cell simulations (see Methods).

**Effect of density modulation.** To highlight the effect of density modulation, we compare the results from the simulation cases of a uniform flat Carbon target and a DM Carbon target in Fig. 2. In both simulation cases, a 66 fs 10 PW laser pulse with a focused intensity of $2 \times 10^{22}$ W/cm² is employed. The upper-left panel illustrates that in the uniform flat target the radiation pressure of a Gaussian laser pulse will expel ions away from the center high-field area as observed previously[18,25,26]. Subsequently, the accelerated Carbon ions are widely divergent and form an exponentially decaying energy spectrum in the lower-left panel. In contrast, the upper-middle panel illustrates that the DM Carbon target with thickness up to 5.5 µm is nearly wholly accelerated after the laser irradiation. Consequently, the accelerated Carbon ions are roughly collimated and form a weakly peaked energy spectrum in the lower-middle panel. The peak energy is roughly equal to 390 MeV ($\Xi_0 \sim 0.0225$), which is in good quantitative agreement with the theoretical prediction. This peaked energy spectrum indicates that roughly uniform hole boring is achieved in the DM target. In addition, the energy conversion efficiency from laser to ions in this case is as high as 27.1%, which is higher than that of the uniform flat target (20.7%).

Besides fast ions, however, a great amount of hot electrons will be generated via collisionless heating mechanisms[36]. During the interaction of the laser pulse with the DM target, these hot electrons are confined by the Coulomb potential of the ionic center. Because the target thickness here (5.5 μm) is much smaller than $c\tau_L$ (20 μm), some hot electrons may travel through the target forward and backward many times. Consequently, the whole target is rapidly heated by such a refluxing of hot electrons. The heating of the target leads to a high thermal pressure, which is harmful to the hole-boring RPA that relies on a negligible thermal pressure. Meanwhile, other mechanisms such as the TNSA, BOA or Coulomb explosion may take part in the torch relay of accelerating ions as the target is heated[11-15]. As a result, some Carbon ions can be accelerated to energies as high as 1 GeV, far beyond the energy predicted by the hole-boring RPA model. However, the energy spread of ion beam appears extremely broad, which greatly limits its usability.

**Effects of substrate and target composition.** To improve the quality of the ion beam, we introduce two further optimizations which can stabilize the hole-boring RPA. Firstly, the DM Carbon target is replaced by a DM two-ion-species (such as $CH_2$) target. Secondly, the DM target is surrounded with a substrate of sufficient thickness. With this setup, a dense block of energetic Carbon ions are generated from the DM $CH_2$ target as highlighted in the upper-right panel of Fig. 2. This dense block of energetic ions represents a well-collimated quasi-monoenergetic ion beam. The lower-right panel of Fig. 2 shows that this ion beam has an opening angle ~7.1° and a FWHM energy spread ~ 28.5% at about 390 MeV. Merely in terms of energy spread, this ion beam seems trifling compared to those generated in the light-sail RPA with ultrathin foils[21-24], collisionless shock acceleration[15], or hole-boring RPA[32] with gas targets. However, considering that this ion beam has a thickness of a few microns together with a high density as solids, the energy spread of 28.5% is fairly satisfactory. For larger numbers of ions, it becomes increasingly difficult to make them move coherently due to the stronger effects of Coulomb repulsion. This might be the reason for the relatively broad energy spread (~ 28.5%) presented here. Fortunately, we find that the beam quality can be well maintained even after the laser irradiation as shown in the Supplementary Movie S1, which displays the whole accelerating process. The proton density and energy-angle distributions are similar to those of Carbon ions and are shown in the Supplementary Fig. S1. Since the protons are preferentially accelerated by other acceleration mechanisms due to their higher charge-to-mass ratio, they have a slightly broader energy spread than the Carbon ions.

In Fig. 2a, the energetic ion blocks display FWHM dimensions of $L_x$~9.2 and ~2.5 μm for cases without and with a substrate, respectively. Meanwhile, the corresponding energy spectra in Fig. 2b are weakly-peaked and highly-peaked, respectively. These results indicate that the substrate plays a crucial role in sustaining the high-quality hole-boring RPA, despite the fact that it does not directly interact with the laser pulse. This is because the return current of cold electrons from the substrate part can suppress the refluxing of hot electrons into the DM part. Fig. 3a indicates that in the DM region the return electrons from the substrate are as dense as the electrons initially from the DM region at $t =$ 33 fs. More details on the runaway of heated DM electrons and the return of substrate electrons in the whole process can be seen in the Supplementary Movie S2. Consequently, Fig. 3b highlights that the plasma temperature has been reduced by a factor of about 10 inside the target and nearly halved at the laser-target interface. Therefore, the ions are mainly accelerated via the hole-boring RPA that assumes a relatively cold target, and other ion acceleration mechanisms, by virtue of the fast electrons, are suppressed.

Without the substrate, however, the peak plasma temperature is as high as 40 MeV in Fig. 3b. The plasma temperature $T_e$ is defined by the averaged kinetic energy of electrons in the center-of-momentum frame. The detailed calculation of $T_e$ in PIC simulations is explained in the Methods section. Considering the peak density ~ $100n_c$, we estimate that the plasma pressure $P_e = kn_eT_e$ could be as high as 6.5 Tbar, where $k$ is the Boltzmann constant. This pressure is already about half of the laser radiation pressure $2I/c \approx 13$ Tbar, therefore, it severely disturbs the hole-boring RPA and lowers the quality of the accelerated ion beam.

In experiments, a substrate will naturally be present to support the DM target in most cases. However, the role of the substrate here is mainly to suppress the refluxing of hot electrons and to cool the DM target. Therefore, there is a critical requirement on the thickness of the substrate. In the case of a thin (3 μm) substrate, the quality of generated ion beam is nearly as poor as that without any substrate (see Supplementary Fig. S2). By a series of simulations, we find that the quality of the generated ion beam firstly improves with the increasing substrate thickness and becomes stable after the substrate is thicker than $c\tau_L$ ~ 20 μm. This implies that a substrate as thick as $c\tau_L$ is sufficient to suppress the refluxing of hot electrons.

To reveal the effect of target composition, we have also performed a particle-in-cell simulation using a DM target of pure Carbons with a substrate. The resulting energy spectrum of Carbon ions has a broader energy spread than that from the DM $CH_2$ target with a substrate in the lower-right panel of Fig. 2. It verifies the advantage of using two-ion-species (such as $CH_2$) targets, in which the temporal oscillation of the longitudinal electric field in the hole-boring RPA is effectively suppressed[37].

**Ion energy scaling.** As suggested in Ref. [38], by lowering the target density according to equation (1), a laser pulse of relatively low intensity can accelerate ions to energies equivalent to that produced by an intense beam at higher density. Figure 4 shows that the peak energy ~ 390 MeV is reproduced in a DM target with $n_0 = 25n_c$ and $L_x = 3$ μm at the intensity $I_0 = 5 \times 10^{21}$ Wcm$^{-2}$ ($a = 42.5$), which is already available nowadays in some laboratories. However, we notice that this is achieved at the cost of broadening the ion energy spread. At a given laser intensity, the peak energy of ions can be increased by lowering $n_0$ and enlarging $L_x$ simultaneously according to equation (2). For instance, the peak energy of 608 MeV and 858 MeV are predicted in DM targets with $n_0 = 60n_c$, $L_x = 3.87$μm and $n_0 = 40n_c$, $L_x = 4.74$μm, respectively. These predictions agree well with the simulation results in Fig. 4. However, the energy spectrum is nearly flat in a broad region for the case of $n_0 = 40n_c$, $L_x = 4.74$μm. If the density further decreases, the target may become relativistically transparent. In this so-called relativistic transparency or near-critical-density regime, the ions may be accelerated up to higher energies by other mechanisms[12,13,39], which is beyond the scope of this paper.

**Discussion**

On the basis of the above results, a route towards the generation of dense blocks of energetic ions has been provided. This route is based on the uniform and efficient hole-boring RPA that occurs when a multi-petawatt Gaussian pulse irradiates a DM target. Our study has highlighted that the refluxing of hot electrons in the DM target can interestingly be suppressed by simply adding a substrate, so the DM target can be kept relatively cold over the whole duration of the hole-boring RPA. Consequently, both the yield and the quality of the accelerated ion beam have been significantly enhanced with our route.

We point out that Coulomb collisions between the accelerated ions and the particles of the substrate have not been considered in our model for particle-in-cell simulations. On one hand, the substrate may obviously slow down the accelerated ions via Coulomb collisions if it is too thick. On the other hand, the substrate should be thick enough to suppress the refluxing of hot electrons. So it would be crucial to control the thickness of the substrate. As mentioned above, a substrate as thick as $c\tau_L \sim 20$ μm is sufficient to suppress the refluxing of hot electrons. This thickness is about three orders of magnitude smaller than the stopping range of these Carbon ions[40], so the effect of Coulomb collisions can be neglected with such thick substrates.

The dense block of well-collimated quasi-monoenergetic ions produced with our route is interesting for the creation of astrophysical-like high energy density conditions[8,41], as well as high pressure materials science[42]. More importantly, it may be applied in the fast ignition of inertial confinement fusion due to the presence of the Bragg peak for monoenergetic ion beams[1,2,6,28,43,44]. As the ion charge density reaches $10^4$ Coulomb/cm$^3$, the current and current density of this energetic ion block is about 50 MA and $10^{14}$ A/cm$^2$, respectively. Such high currents and current densities more than meet the requirement for the fast ignition of laser fusion or heavy-ion fusion. The Carbon ion energy of 390 MeV is also close to the optimized ion energy per nucleus for fast ignition[28,44].

Note that the on-axis areal density and energy fluence of the ion beam produced with our route are as high as 0.1 mg/cm$^2$ and 0.2 GJ/cm$^2$, respectively. To the best of our knowledge, both of them are much higher than those achieved by other laser-driven ion acceleration schemes. The total number of energetic ions reaches $2\times10^{12}$, which is even comparable to the particle number per bunch to be produced by the future heavy-ion synchrotron, SIS 100, under construction at GSI[45,46]. However, here the duration of ion beam is under 100 fs, compared to that from SIS 100 typically over 20 ns. Therefore, such a unique ion beam of ultrashort duration and ultrahigh areal density are particularly suitable for isochoric heating of solid-density matter[47]. With the ion beam produced by our route, isochoric heating to the keV level would be feasible.

We admit that this study is still at the stage of theoretic research and the main challenge in realizing such experiments may come from the fabrication of such DM targets. Owing to the continuous advance of micro/nano-fabrication technology, DM targets may be achieved by some potential approaches, such as the pulsed laser deposition technique[34], or layer-by-layer nanoarchitectonics[35]. In fact, density-tunable foam-based multi-layered targets have been obtained by the pulsed laser deposition technique, and employed in recent ion acceleration experiments[48]. Using pulsed laser deposition, porous nanostructured Carbon films can be superimposed on the surface of a substrate to grow a foam layer. By tuning the parameters (gas pressure, fluence, flow direction, target-to-substrate distance etc.), foam layers with the density and thickness in the range of 1-1000 mg/cm$^3$ and 5-80 micrometers, respectively, have already been achieved[34,48]. If the parameters can be dynamically tuned during the deposition, a DM foam layer could conceivably be formed on the surface of a substrate as described in our proposal. With respect to the layer-by-layer nanoarchitectonics, Carbon nanotube films have been structured on the nanometer scale (average 100 nm gaps between 10-20 nm bundles) to provide a uniform near-critical-density plasma at the front of an ultrathin foil[49]. If one could further control the gaps between the bundles of Carbon nanotubes, a DM target would then become feasible. In addition, gas targets with modulated density profiles could be an alternative target configuration in the demonstration of our proposed scheme. It is worth noting that a high-quality ion beam has been demonstrated in the interaction between an intense infrared laser pulse and a gas target with an approximately triangular density profile instead of a uniform density profile[32]. If the density profile of the gas target could be modulated further, the proposed uniform laser hole boring and consequent high-quality ion acceleration might also be achieved by infrared lasers at a greatly reduced laser intensity.

**Methods**

**Numerical simulations.** The generation of dense blocks of energetic ions by multi-petawatt laser pulses is studied with a self-implemented particle-in-cell code[33,37,39]. 2D3V (two dimensions in space and three dimensions in velocity space) particle-in-cell simulations are performed using a box spanning in $-10 \leq x/\lambda \leq 40$ and $-30 \leq y/\lambda \leq 30$. The simulation box consists of 10000×9600 cells, and 240 macroparticles per cell are allocated in the plasma region. In simulation, a laser pulse propagate in the positive $x$ direction along $y = 0$. For reference, the peak of the laser pulse is assumed to arrive at the focus surface $x = 0$ at $t = 0$ if there is no plasma. All simulations begin at $t = -50T_0$ and end at $t = 70T_0$, where $T_0 \approx 3.3$ fs is the wave period of the laser.

In the first set of simulations, a comparison is made between four different target configurations. The first case uses a uniform flat Carbon target with $\rho \approx 0.33$ g/cm$^3$ (the electron number density $n_e = 87.5n_c$) that occupies in $-2.75 \leq x/\lambda \leq 2.75$ and $-25 \leq y/\lambda \leq 25$. The second case uses a DM Carbon target that has $\rho = \rho_0 \exp(-2.77x^2/L_x^2)\exp(-r^2/\sigma^2)$ with the FWHM dimension $L_x = 3$ μm and peak density $\rho_0 \approx 0.33$ g/cm$^3$ (the maximum electron number density $n_0 = 87.5n_c$). For simplicity, this DM Carbon target is cut off at $\rho = \rho_0/10$ and hence the on-axis thickness is about 5.5 μm. The third case uses a DM CH$_2$ target that is composed of Carbon ions and protons in 1:2 number ratio. The mass density profile of this target is the same as that in the second case, but here the maximum electron number density $n_0 = 100n_c$. Further, this DM CH$_2$ target is surrounded by a uniform substrate with $n_e = 100n_c$ and $m_i/q_i = 3672m_e/e$ in the region $0 \leq x/\lambda \leq 30$ and $-25 \leq y/\lambda \leq 25$. The fourth case uses a DM Carbon target with a substrate. The DM Carbon target is the same as that in the second case, while the substrate is the same as that in the third case. In each case, a circularly-polarized Gaussian laser pulse of peak intensity $I_0 = 2 \times 10^{22}$ Wcm$^{-2}$ ($a \approx 85$), duration $\tau_L = 20T_0$ and spot size $\sigma = 4\lambda$ is employed, where $\lambda = 1\,\mu$m is the laser wavelength. Such an intense pulse could be achieved by the tight focusing of a 10 PW 0.7 kJ laser beam, which is within the designed capacity of a number of high-power laser facilities, such as ELI, LFEX, OMEGA EP, and HiPER[50]. In both the DM Carbon target and the DM CH$_2$ target, the total number of Carbon atoms is as high as $2 \times 10^{12}$. In the 2$^{nd}$- 4$^{th}$ cases, a uniform hole-boring velocity $v_b \sim 0.13c$ ($\sqrt{\Xi_0} \sim 0.15$) is predicted by equations (1) and (2).

In the second set of simulations, we validate the scaling law of the peak energy of accelerated ions. Simulations are performed with the following combinations of laser peak intensities, target peak densities and target dimensions: (1) $a = 42.5$, $n_0 = 25n_c$, $L_x = 3$ μm; (2) $a = 85$, $n_0 = 100n_c$, $L_x = 3$ μm; (3) $a = 85$, $n_0 = 60n_c$, $L_x = 3.87$ μm; and

(4) $a = 85$, $n_0 = 40n_c$, $L_x = 4.74$ μm. In all four of these simulations, the remaining laser-target parameters are set equal to those of the third simulation in the first set.

**Calculation of temperature.** We define the plasma temperature as $T_e = 2\langle E^*_{kin}\rangle/3k$, where $\langle E^*_{kin}\rangle$ is the averaged kinetic energy of electrons in the center-of-momentum (COM) frame. We assume that the velocities of all macroparticles in the laboratory frame have already been obtained in PIC simulation. Then, the velocity of the electron mass center $\langle \mathbf{V}\rangle$ can be calculated as

$$\langle \mathbf{V}\rangle = \frac{\sum_{n=1}^{N} \gamma_n \mathbf{V}_n}{\sum_{n=1}^{N} \gamma_n}, \qquad (3)$$

where N is the macroparticle number for electrons in the considered cell, and $\mathbf{V}_n$ and $\gamma_n = (1 - |\mathbf{V}_n/c|^2)^{-1/2}$ are the velocity and Lorentz factor of the *n*-th macroparticle, respectively. Then the Lorentz factor of the *n*-th macroparticle in the COM frame can be obtained by the Lorentz transformation as

$$\gamma^*_n = \gamma_n \langle \gamma\rangle \left[1 - \frac{\langle \mathbf{V}\rangle \mathbf{V}_n}{c^2}\right], \qquad (4)$$

where $\langle \gamma\rangle = (1 - |\langle \mathbf{V}\rangle/c|^2)^{-1/2}$ is the Lorentz factor related to the velocity of the electron mass center $\langle \mathbf{V}\rangle$. Consequently, the averaged kinetic energy of electrons in the COM frame can be calculated from

$$\langle E^*_{kin}\rangle = m_e c^2 \sum_{n=1}^{N} (\gamma^*_n - 1)/N. \qquad (5)$$

Using Eqs. (3) and (4), the above equation can be finally simplified to

$$\langle E^*_{kin}\rangle = m_e c^2 \left(\frac{\sum_{n=1}^{N} \gamma_n}{N\langle\gamma\rangle} - 1\right). \qquad (6)$$

Acknowledgments

This work was supported in part by the National Basic Research Program of China (Grant No. 2013CBA01504) and the National Natural Science Foundation of China (Grant Nos. 11405108, 11421064, 11129503 and 11374210). S.M.W. and M.C. appreciate the supports from National 1000 Youth Talent Project of China. Z.M.S acknowledges the support of a Leverhulme Trust Research Project Grant at the University of Strathclyde. Simulations have been carried out at the PI cluster of Shanghai Jiao Tong University. We appreciate the help by Thomas Wilson for improving the English writing.

Additional information.

Supplementary information accompanies this paper at http://www.nature.com/srep

Competing financial interests: The authors declare no competing financial interests.

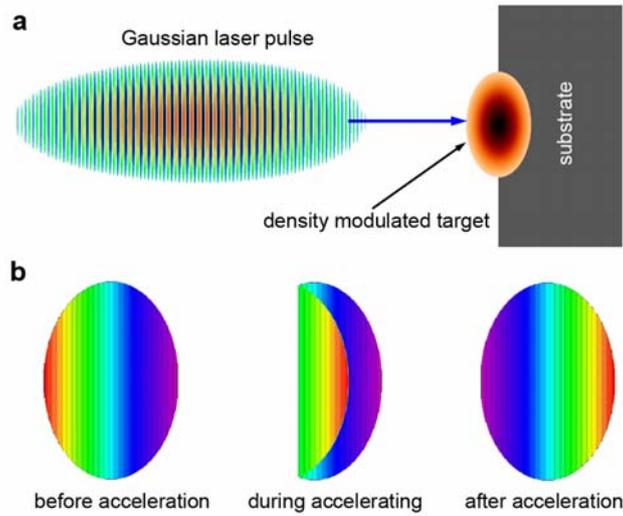

**Figure 1 | Schematic of a uniform hole-boring acceleration.** (a) A density modulated (DM) target is half-embedded in a conductive substrate, and a circularly-polarized laser pulse irradiates the target from the left side. (b) The ions are accelerated layer by layer from left to right. Finally all layers of the ions are accelerated to a roughly uniform speed but in the reverse order. The initial sequence numbers of the layers are represented by colors in the order of a rainbow.

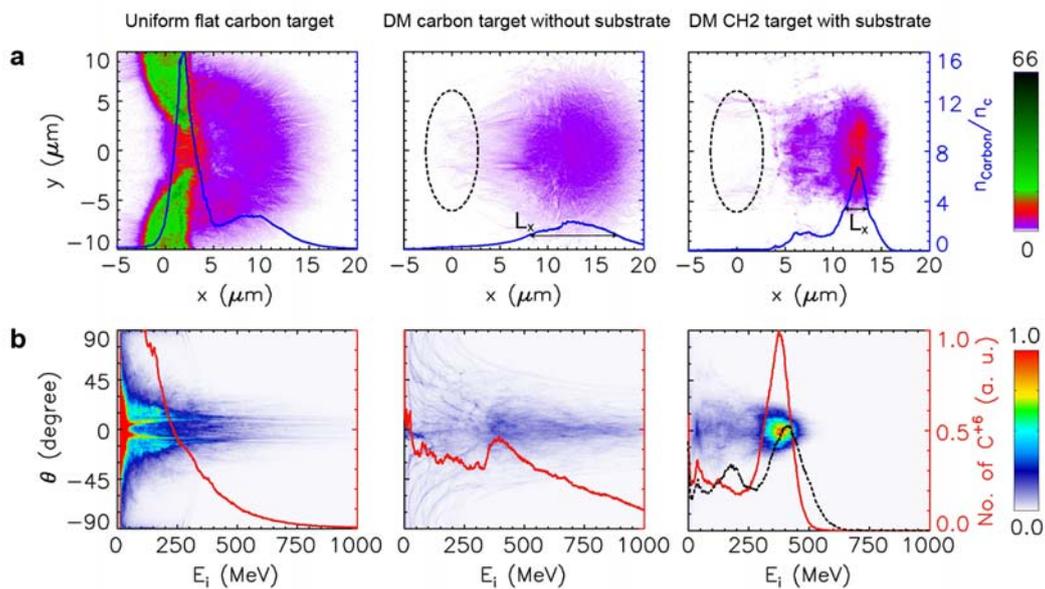

**Figure 2 | Comparison between different target configurations.** (a) Carbon ion distributions (color contour) and densities averaged over $|y| \leq 4\,\mu m$ (blue line) at $t =$ 165 fs obtained from the simulation cases using the uniform flat Carbon target, DM Carbon target without substrate and DM $CH_2$ target with a substrate, respectively, for

the panels from left to right. Each DM target is initially located in the dashed ellipse, and $L_x$ indicates the instantaneous FWHM dimension in $x$ direction. In all cases, a circularly-polarized Gaussian laser pulse of peak intensity $I_0 = 2 \times 10^{22}$ Wcm$^{-2}$ ($a \approx 85$), duration $\tau_L = 66$ fs and spot size $\sigma = 4\,\mu$m is employed. (b) The corresponding ion energy-angle distributions (color contour) and energy spectra (red line). The spectrum in the case of a DM Carbon target with a substrate (black dashed line) is also drawn for comparison in the lower-right panel.

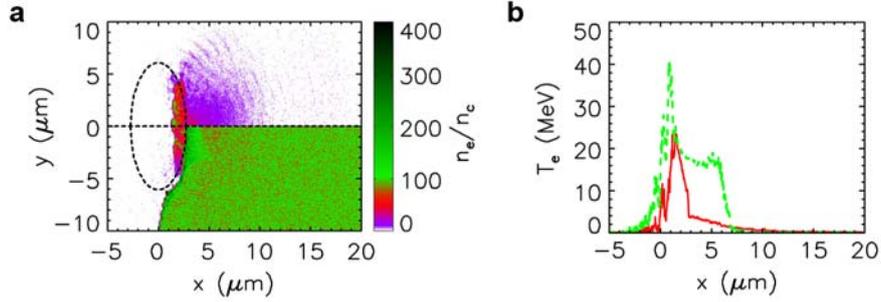

**Figure 3 | Suppression of plasma heating by a substrate.** (a) Distributions of the electrons from the DM part (upper half) and the electrons from the substrate part (lower half) at $t = 33$ fs in the simulation case using the DM CH$_2$ target with a substrate. Both are symmetric about the $x$-axis, and the dashed ellipse demarcates the initial boundary of the DM CH$_2$ target. (b) Plasma temperatures averaged over $|y| \leq 4\,\mu$m for the cases with (red solid line) and without (green dashed line) a substrate.

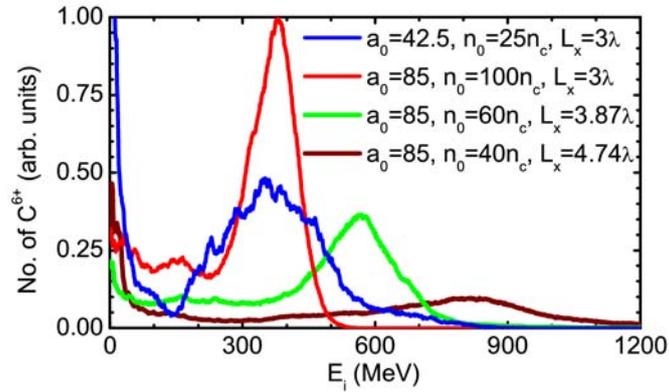

**Figure 4 | Ion energy scaling.** Carbon ion energy spectra obtained from the interactions of laser pulses at different intensities $a_0$ with DM CH$_2$ targets of different peak densities $n_0$ and FWHM dimensions $L_x$ in $x$ direction.